\documentclass{emulateapj}

\shorttitle{Supersonic Evershed flow outside Sunspots}
\shortauthors{Mart\'\i nez Pillet et al.}

\begin{document}

\title{Supersonic continuation of the Evershed flow outside a Sunspot as observed with Hinode
}

\author{V. Mart\'\i nez Pillet}
\affil{Instituto de Astrof\'\i sica de Canarias,
    La Laguna, 38200, Spain}
\email{vmp@iac.es}

\author{Y. Katsukawa}
\affil{National Astronomical Observatories of Japan, 2-21-1, Osawa, Mitaka, Tokyo, 181-8588, Japan}

\and

\author{K. G. Puschmann and B. Ruiz Cobo\altaffilmark{1}}
\affil{Instituto de Astrof\'\i sica de Canarias, La Laguna, 38200, Spain}

\altaffiltext{1}{University of La Laguna, La Laguna, Spain.}

\begin{abstract}
We report on the discovery of mostly horizontal field channels just outside
sunspot penumbrae (in the so-called `moat' region) that are seen to sustain
supersonic flows (line-of-sight component of 6 km s$^{-1}$).  The spectral
signature of these supersonic flows corresponds to circular polarization
profiles with an additional, satellite, third lobe of the same sign as the
parent sunspot' Stokes $V$ blue lobe, for both downflows and upflows.  This is
consistent with an outward directed flow that we interpret as the continuation
of the magnetized Evershed flow outside sunspots at supersonic speeds.  In
Stokes $Q$ and $U$, a clear signature of a transverse field connecting the two
flow streams is observed. Such an easily detectable spectral signature should
allow for a clear identification of these horizontal field channels in other
spectropolarimetric sunspot data. For the spot analyzed in this paper, a
total of 5 channels with this spectral signature have been unambiguously found.    
\end{abstract}



\section{Introduction}

The supersonic\footnote{We use the term supersonic in a broad sense meaning
velocities higher than 6 km s$^{-1}$.} character of the Evershed flow inside
sunspot penumbrae is well established \citep{wiehr95,deltoro01,bellot04}.
Already, \citet{wiehr95} measured velocities (from Stokes $I$ bisectors) larger
than 5 km s$^{-1}$. More recent spectropolarimetric measurements have clearly
established the existence of ubiquitous supersonic flows everywhere in the
penumbra.  \ion{Fe}{1} 1.5$\mu$m infrared polarization spectra are particularly
well suited for this purpose as the strong Zeeman sensitivity of these lines,
and to some extent the associated lower formation heights, allow to separate
the sub- and supersonic components \citep{deltoro01,bellot04}. Using Hinode/SP
data (in the 6301.5 and 6302.5 \AA~\ion{Fe}{1} line pair) \citet{bellot07}
identified Stokes $V$ signals originated in the so-called penumbral dark cores
\citep{scharmer02} that display three lobe profiles with velocities in the
range of 6-7 km s$^{-1}$. With similar Hinode data, speeds of up to 9.5 km
s$^{-1}$ have been reported recently \citep{bellot09}.  For these profiles to
show up, it is crucial that the spot displays a sizeable line-of-sight (LOS)
component of the Evershed flow as helped by off disk center observations. These
highly Doppler shifted signals observed in both infrared and visible data
clearly point to a magnetized nature of the Evershed flow, which is compatible
with the uncombed penumbral model of \citet{solanki93}. In this model, nearly
horizontal flux tubes with weaker field strengths \citep{martinezpillet00}
carry the Evershed flow. The horizontal tubes are embedded in a more vertical
background field that has comparatively speaking smaller velocity flows. The
dark core penumbral filaments have been solidly associated with the
observational signature of horizontal penumbral flux tubes
\citep{borrero07,ruizcobo08}. The alternative proposal of a field free ambient
where the Evershed flow is localized \citep{scharmer06} faces the serious
problem that no Doppler-shifted Stokes V signals are generated in this model.

\begin{figure*}
\centerline{
\epsscale{1.0}
\plotone{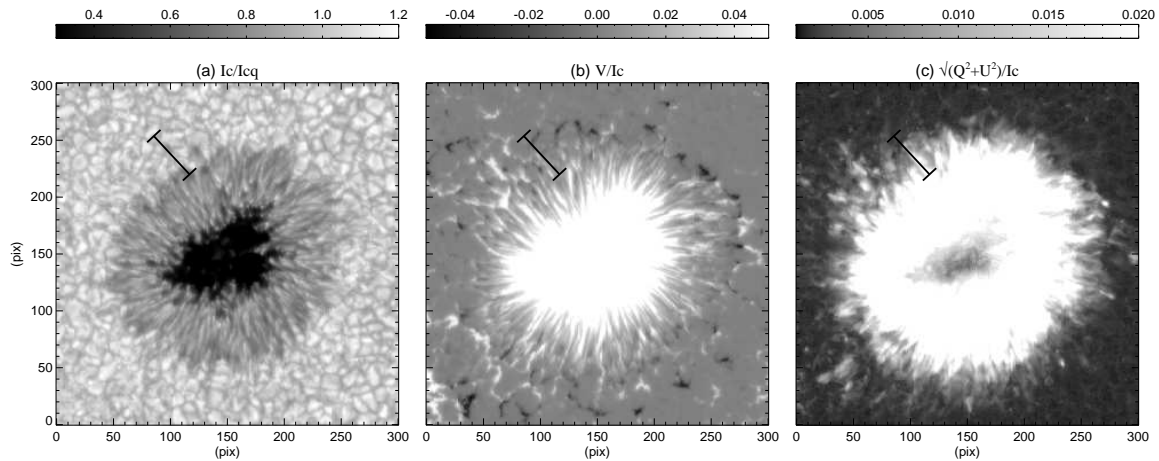}
}
\caption{Hinode/SP slit reconstructed continuum intensity map, circular polarization 
and total linear polarization magnetograms (from left to right). 
The black segment indicates the location of the  
enhanced transverse field region analyzed in this paper.
\label{fig1}}
\end{figure*}

While the scenario in which  penumbral flows take place has been clarified in
recent years, the origin of the Evershed flow and its final fate remain
unclear.  According to the simulations of \citet{sch98}, the evolution of
buoyant flux tubes inside the penumbra is such that an internal flow develops
and attains speeds of up to 14 km s$^{-1}$.  These high speed flows could propagate
beyond the outer penumbral boundary in some of these simulations
\citep[see][]{sch02}. In contrast, \citet{westendorp97} found clear evidence
that a large fraction of the flow is returning into the photosphere at the
penumbral boundary, a well established result confirmed by a large number of
modern observations. While this explains the signature of an abrupt end of the
flow, it still does not answer the question of what happens to it once
it dives down into the photosphere and outside of the main body of the spot.  Two
recent results are relevant to understand what could be going on. First of all,
the sunspot surroundings (the `moat') have been seen to harbour horizontal
filamentary magnetic fields as detected in SOHO/MDI magnetograms averaged over 10
hours \citep{sainz05}. The bipolar Moving Magnetic Feature (MMF) activity is
seen to start well inside the penumbra and propagate in a coherent manner
away from the spot \citep[][who detected a sea-serpent
configuration]{sainz05,ravindra06,cabrera06,sainz08}.  A second observational
result of importance to ascertain the further evolution of the Evershed flow is
the close relation found by the existence of radially oriented penumbral
filaments and the moat flow \citep{vargas07}. This has led these authors to
propose a link between the Evershed flow and the moat flow. All in all, the
subsequent evolution of the flow beyond the penumbral border is unclear. In
particular, it is not known whether it maintains the supersonic character that
both observations and simulations suggest. 

Strong (supersonic) downflows have been detected by Hinode in the moat region
and, to a smaller extent, in other solar regions \citep{shimizu08}.  Similar
large upflows have only been detected in the quiet sun by \citet{socas05}, who
proposed that they represent evidence of a failed convective collapse process.
All these flows have a clear spectral signature in Stokes $V$ profiles that
display a strong Doppler-shifted third satellite lobe. In this Letter, we
report evidence of {\em horizontal magnetic links} between supersonic
downflows and upflows at and beyond the penumbral boundary. It is proposed that
these signatures reveal sporadic continuations of the (supersonic) Evershed
flow outside the spot.

\section{Observations and data selection}

In order to understand to what degree the moat region is pervaded by horizontal
magnetic fields, Hinode/SP observations \citep{kosugi07,tsuneta08} of a sunspot
located very close to disk center were selected. It is important for this study
that the moat region was located near disk center to maximize the sensitivity
to horizontal fields in Stokes $Q$ and $U$.  Disk center observations also
allow to sense deeper into the photosphere, something which we consider
important to improve the visibility of any horizontal magnetic fields residing
in the moat. The Hinode/SP map was obtained on the 28th of February of 2007 at
18:25 UT in the main, positive polarity, sunspot of NOAA 10944, which was
located only 1.2 degrees off disk center.  This data set was included in the
analysis of \citet{ichimoto08}.

\begin{figure*}
\centerline{
\epsscale{1.0}
\plotone{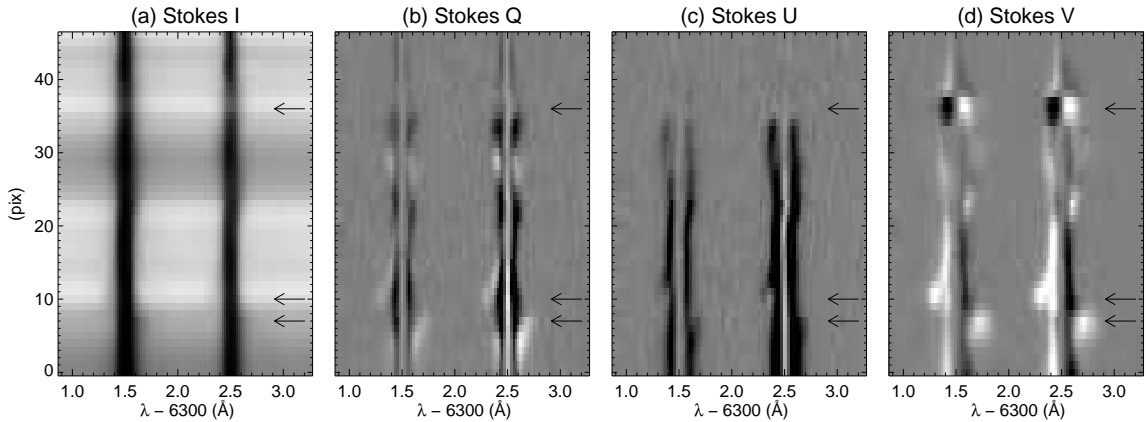}
}
\caption{From left to right: Stokes $I$, $Q$, $U$ and $V$ profiles along the transverse
field region indicated in Fig. \ref{fig1}. The spectral signature of the supersonic
downflows and upflows is marked with two horizontal arrows (coordinates \#7 \& \#10)
as well as the network point used as reference (coordinate \#36).
\label{fig2}}
\end{figure*}

Figure \ref{fig1} shows slit reconstructed images of the continuum intensity,
Stokes $V$ magnetogram\footnote{As computed by the routine stksimages\_
sbsp.pro of the Hinode solarsoft package.} (scaled to display longitudinal
fields over the outer penumbra and the moat) and a total linear polarization
map (scaled to provide good visibility of transverse fields in the moat
region).  Far-wing Stokes $V$ magnetograms at about 250 m\AA~at each side of
line center were also constructed to help identifying strong up/downflow
regions in the moat \citep[following][]{ichimoto07,shimizu08}. The linear
polarization map was used as a reference for the existence of filamentary field
aggregations beyond the penumbra. In this map, a total of five regions were
identified as transverse field concentrations that coincide along their
length with flow signatures in the far-wing magnetograms (although a
thorough study was not performed, in particular many more transverse extensions
are easily seen in the LP map of Fig. \ref{fig1}).  In contrast to the results
of \citet{shimizu08}, who detected only downflows, the use of the transverse
fields as a reference has allowed the identification of both upflows and
downflows in the moat.  The experience gained from the use of the far-wing
magnetograms showed that the upflows observed have a tendency to be slightly
weaker than the downflows, but the example presented in this Letter has both
flows of a similar magnitude.

All of the five regions selected with transverse fields in the moat display a
downflow/upflow spectral signature that we identify with a supersonic extension
of the Evershed flow along mostly horizontal field lines outside the spot.  In
this Letter, we present a case example of this spectral signature and offer a
simplified analysis of their quantitative implications. A more in depth study
of all the regions in this map, including two-components SIR inversions
\citep{sir92}, is postponed to a future paper. The location of the case
selected for this work as representative of the spectral signature of the
continuation of the Evershed flow in the moat region is marked by a sector in
Fig. \ref{fig1}.  As it is evident in the linear polarization frame of Fig.
\ref{fig1}, this region is coexisting with transverse fields in the moat but is
not particularly prominent in the longitudinal magnetogram (that mostly shows
the well-known MMF activity).  The Stokes profiles along the segment in Fig.
\ref{fig1} are shown in Figure \ref{fig2} (note that this is a composite that
does not corresponds to any actual slit position). This figure already displays
the fingerprint of the Evershed flow continuation which is marked by the two
arrows near the bottom of the frames. The Stokes $V$ profile shows the
existence of a satellite third lobe at the red and blue sides of the rest
position. The downflows and upflows are spatially separated by slightly less
than 1 arcsec (the scale in this figure is $\sim$ 0.15 arcsec/px).
Interestingly, the flows are also evident in the Stokes $Q$ and $U$ signals.
The first downflow occurs within the penumbra \citep[and it would correspond to
a supersonic version of the cases found by][]{westendorp97}. The upflow occurs
just outside the visible penumbra. Continuous $Q$ and $U$ signals show that we
are tracing an extension of horizontal field lines from within the penumbra. It
is important to point out that these horizontal fields must reside at the line
forming region but probably slightly above the $\tau =1$ layer as no darkening
is seen in the continuum maps. This changes at coordinate \#23 (ref. to Fig.
\ref{fig2}), where the penumbra reappears in continuum. A weaker downflow
occurs at this location (not marked) and has a polarity opposite to that of the
spot.  The region where the penumbra is seen to reappear coincides also with
weaker $Q$, $U$ and $V$ signals indicating that the field lines are probably
buried in deeper layers, filling more the continuum forming layer than the line
forming region. 
Similar (albeit slightly less evident) penumbral reappearances 
are seen in two of the other four cases of flow extensions detected in 
this sunspot. 

\section{Supersonic flows signatures}

\begin{figure}
\epsscale{1.0}
\plotone{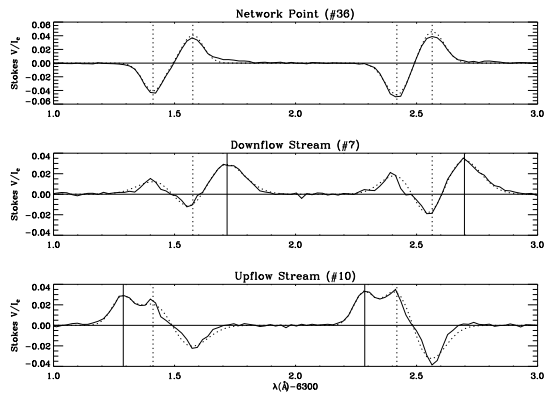}
\caption{Top panel: \ion{Fe}{1} 6301 and 6302 profiles (solid lines)
from the network point used as wavelength reference and the corresponding
2-gaussian fit (dotted line). The centers of the two gaussians for each lobe
are given by the vertical dotted lines. Mid panel: Profiles from the downstream
point and 3-gaussian fit.  The vertical dotted lines indicate the center of the
red lobe of the network point whereas the vertical solid lines mark the centers
of the observed red shifted satellite lobes. Bottom panel: the same as above
but for the upstream point. The vertical dotted lines indicate now the center
of the blue side lobes of the network point.
\label{fig3}}
\end{figure}

In order to quantify the flow velocities that are needed to reproduce the
spectral satellites seen in Fig. \ref{fig2}, a simplified analysis was made
\citep[following the multilobe analysis made
by][]{shimizu08}.  It is based on fitting gaussians to the various lobes
observed in Stokes $V$ and on referring their local maximums to a wavelength
`rest' position. For this reference, we select the network point at coordinate
\#36 as it is well-known that these concentrations do not display large
velocities \citep[see] [for the \ion{Fe}{1} lines used in this
work]{martinezpillet97}.  Two gaussians are first fitted to the network point
obtaining the rest wavelengths for both the blue and red lobes. Then, a three
gaussian combination is fitted to the profiles at coordinates \#7 (downflow)
and \#10 (upflow). The resulting fits are shown in Figure \ref{fig3}. The
maximum of the extreme red and blue lobes, once compared with the center of the
corresponding network lobes, provide an estimate of the flow speed that must be
present to produce these satellite spectral signatures. In the case of the
downflow of point \#7, the results are 6.7 km s$^{-1}$ for \ion{Fe}{1} 6301.5
\AA~ and of 6.3 km s$^{-1}$ for \ion{Fe}{1} 6302.5 \AA. In the upflowing stream
of point \#10, we obtain -5.9 km s$^{-1}$ and -6.3 km s$^{-1}$, for the same
lines respectively. These velocities correspond to LOS components of the stream
flow. As the flow signature is also seen in Stokes $Q$ and $U$, it is clear
that there is a significant inclination of the corresponding field lines
indicating that the flow speeds could be considerably larger than 6 km
s$^{-1}$. A four gaussian fit to the two non-network profiles in Fig.
\ref{fig3} was also attempted, but no convincing results were obtained. This
indicates that the Stokes $V$ profile corresponding to the satellite profiles
are well reproduced with only one lobe, something that indicates the presence
of strong gradients in this component. While we postpone a more detailed
discussion of these profiles to a future paper, we note that already the
supersonic profiles observed by \citet{socas05} were explained with a
single-lobe Doppler shifted component with very strong LOS gradients.
     
As it can be seen in Fig. \ref{fig3}, the two blue and red satellite lobes are
positive (which corresponds with the white signals at each side of the rest
Stokes $V$ profile in Fig. \ref{fig2}). It is important to understand the
implications of these signs in the Doppler-shifted component as they readily
show that the supersonic flow corresponds to an {\it outward} directed flow from the
spot (as it should be for an extension of the Evershed flow outside the spot).
The rationale is as follows. The parent sunspot positive polarity produces a
Stokes $V$ profile with a positive blue lobe and a negative red lobe (see Fig.
\ref{fig2}). If these field lines have to return back into the photosphere, the
polarity is reversed (positive red lobe). If, in addition, there is a redshift
of the signal created by a downflowing stream, the lobe that will be seen
further away from the rest wavelength will, then, be positive. Similarly, if
the field lines reemerge to the surface, they preserve the spot polarity
(positive blue lobe) which together with an upflowing stream (blue shifting the
profile) produces a displaced positive Stokes $V$ lobe.  The same arguments
lead to the conclusion that a radial flow streaming into the spot would have
had negative Doppler shifted lobes in Fig. \ref{fig3} (or black satellites at
both sides of the main profile in Fig. \ref{fig2}).  All five regions
identified in the surroundings of this sunspot with strong transverse fields
have shown positive satellite signals in Stokes $V$.
 
\section{Conclusions}

We report on the finding of highly Doppler-shifted Stokes $V$ satellite lobes
at each side of the normal rest profiles, revealing the presence of downflowing
and upflowing material moving away from the spot. The spectral signature of the
Doppler shifted Stokes $V$ lobes is seen along extensions of the penumbra that
display strong transverse fields. 

The flow speeds inferred by a simple gaussian fit show LOS directed speeds of
$\sim$ 6 km s$^{-1}$. The transverse field character of this signature
(at Hinode/SP resolution)
suggests that the real flow along the field lines would be considerably larger
and is likely to be supersonic. Five such transverse field extensions with positive
satellite Stokes $V$ lobes in the blue and red side of the rest spectral
profile have been detected.  The fact that the sign of the these lobes is
positive (for a positive polarity spot) indicates that the local flow is
directed in the radial direction {\em away} from the spot. This property,
together with the high-speed (supersonic) 
character and the presence of, at least, five such
channels all around the spot, strongly suggest that these profiles represent
the continuation of the Evershed flow outside the spot, into the moat region.

It is also clear that, being the Evershed flow ubiquitous within the penumbra,
the finding of five cases of its extension outside the spot can hardly explain
where the bulk of the penumbral flow ends. We hypothesize that a mechanism such
as convective pumping \citep{thomas02} is able to submerge below the surface
most of the (horizontal) field lines that carry the majority of the Evershed
{\it flow}. However, it is clear that this mechanism will not always be perfect and
often will fail to submerge a particular set of field lines that might become
visible, corresponding to the cases detected in this work. It is also clear
that our findings prove that the Evershed flow can continue beyond the
penumbral boundary preserving the supersonic character it normally attains
within the penumbra. It is unclear at present whether the preservation of this
supersonic character is precisely the reason why these flows are not hidden
below the surface, allowing them to be detected in the Hinode data. It could
well be that if the flow remains supersonic, the convective pumping mechanism
is less efficient at submerging the corresponding field lines.  More systematic
studies of how often one finds these satellite spectral signatures surrounding
sunspots and their temporal evolution are needed to settle this point.

\acknowledgments
Hinode is a Japanese mission developed and launched by
ISAS/JAXA, with NAOJ as domestic partner and NASA and
STFC (UK) as international partners. It is operated by these
agencies in cooperation with ESA and NSC (Norway).
This work has been partly funded through project ESP2006-13030-C06-01.
{\it Facilities:} \facility{Hinode}.



\begin{thebibliography}{}

\bibitem[Bellot Rubio et al.(2004)]{bellot04} Bellot Rubio, L.R., Balthasar, H. \& Collados, M. 2004,
\aap, 427, 319

\bibitem[Bellot Rubio et al.(2007)]{bellot07} Bellot Rubio, L.R., Tsuneta, S., Shimizu, T. et al. 
2007, \apj, 668, L91

\bibitem[Bellot Rubio (2009)]{bellot09} Bellot Rubio, L. private communication

\bibitem[Borrero(2007)]{borrero07} Borrero, J.M. 2007, \aap, 471, 967

\bibitem[Cabrera Solana et al.(2006)]{cabrera06} Cabrera Solana, D., Bellot Rubio, L.R.,
Beck, C., del Toro Iniesta, J.C., 2006, \apj, 649, L41

\bibitem[Ichimoto et al.(2007)]{ichimoto07} Ichimoto, K., Shine, R.A., Lites, B.W. et al.
2007, PASJ, 59, S593 

\bibitem[Ichimoto et al.(2008)]{ichimoto08} Ichimoto, K., Tsuneta, S., Suematsu, Y. et al.
2008, \aap, 481, L9 

\bibitem[Kosugi et al.(2007)]{kosugi07} 
Kosugi, T., Matsuzaki, K., Sakao, T. et al.
	2007, Sol. Phys., 243, 3 

\bibitem[Mart\'\i nez Pillet et al.(1997)]{martinezpillet97} Mart\'\i nez
Pillet, V., Lites, B.W. \& Skumanich, A. 1997, \apj, 474, 810

\bibitem[Mart\'\i nez Pillet(2000)]{martinezpillet00} Mart\'\i nez
Pillet, V., 2000, \aap, 361, 734

\bibitem[Ravindra(2006)]{ravindra06} Ravindra, B. 2006, Sol. Phys., 237, 297

\bibitem[Ruiz Cobo \& del Toro Iniesta(1992)]{sir92} Ruiz Cobo, B. \& del Toro Iniesta, J.C.
1992, \apj, 398, 375

\bibitem[Ruiz Cobo \& Bellot Rubio(2008)]{ruizcobo08} Ruiz Cobo, B. \& Bellot Rubio, L.R. 
2008, \aap, 488, 749

\bibitem[Sainz Dalda \& Bellot Rubio(2008)]{sainz08} Sainz Dalda, A. \& Bellot Rubio
L.R. 2008, \aap, 481, L21

\bibitem[Sainz Dalda \& Mart\'\i nez Pillet(2005)]{sainz05} Sainz Dalda, A. \& Mart\'\i nez
Pillet, V. 2005, \apj, 632, 1176

\bibitem[Scharmer et al.(2002)]{scharmer02} Scharmer, G.B., Gudiksen, G.V., Kiselman, D. et al.
2002, Nature, 420, 151

\bibitem[Scharmer \& Spruit(2006)]{scharmer06} Scharmer, G.B. \& Spruit, H.C. 2006, \aap, 460, 605 

\bibitem[Schlichenmaier et al.(1998)]{sch98} Schlichenmaier, R., Jahn, K. \& Schmidt, H.U. 1998, \aap,
337, 897

\bibitem[Schlichenmaier(2002)]{sch02} Schlichenmaier, R. 2002, AN, 323, 303

\bibitem[Shimizu et al.(2008)]{shimizu08} Shimizu, T., Lites, B.W., Katsukawa, Y. et al.
2008, \apj, 680, 1467

\bibitem[Socas-Navarro \& Manso Sainz(2005)]{socas05} Socas-Navarro, H. \& Manso Sainz, R. 2005,
\apjl, 620, L71

\bibitem[Solanki \& Montavon(1993)]{solanki93} Solanki, S.K. \& Montavon, C.A.P. 1993, \aap, 275, 283

\bibitem[Thomas et al.(2002)]{thomas02} Thomas, J.H., Weiss, N.O., Tobias, S.M., Brummell, N.H. 
2002, Nature, 420, 390

\bibitem[del Toro Iniesta et al.(2001)]{deltoro01} del Toro Iniesta, J.C., Bellot Rubio, L.R.
 \& Collados, M. 2001, \apjl, 549, L139

\bibitem[Tsuneta et al.(2008)]{tsuneta08} 
Tsuneta, S., Ichimoto, K., Katsukawa, Y. et al. 
	2008, Sol. Phys., 249, 167 

\bibitem[Vargas Dom\'\i nguez et al.(2007)]{vargas07} Vargas Dom\'\i nguez, S., Bonet, J.A.,
Mart\'\i nez Pillet, V.  et al. 2007, \apjl, 660, L165

\bibitem[Westendorp Plaza et al.(1997)]{westendorp97} Westendorp Plaza, C., del Toro Iniesta,
J.C., Ruiz Cobo, B. et al. 1997, Nature, 389, 47

\bibitem[Wiehr(1995)]{wiehr95} Wiehr, E. 1995, \aap, 298, L17

\end{thebibliography}
\end{document}